# Strong suppression of the resistivity near the transition to superconductivity in narrow microbridges in external magnetic fields

Xiaofu Zhang,[1] Adriana E. Lita,[2] Konstantin Smirnov,[3,4] HuanLong Liu,[1] Dong Zhu,[1] Varun B. Verma,[2] Sae Woo Nam,[2] and Andreas Schilling[1]

[1]*Department of Physics, University of Zürich, Winterthurerstrasse 190, 8057 Zürich, Switzerland*

[2]*National Institute of Standards and Technology, 325 Broadway, Boulder CO 80305, USA*

[3]*Moscow State Pedagogical University, Malaya Pirogovskaya str. 22, Moscow 127055, Russia*

[4]*National Research University Higher School of Economics, Myasnitskaya str. 20, Moscow 101000, Russia*

*We have investigated a series of superconducting bridges based on homogeneous amorphous WSi and MoSi films, with bridge widths w ranging from 2 μm to 1000 μm and film thicknesses d ~ 4-6 nm and 100 nm. Upon decreasing the bridge widths below the respective Pearl lengths, we observe in all cases distinct changes in the characteristics of the resistive transitions to superconductivity. For each of the films, the resistivity curves R(B,T) separate at a well-defined and field-dependent temperature $T^*(B)$ with decreasing the temperature, resulting in a dramatic suppression of the resistivity and a sharpening of the transitions with decreasing bridge width w. The associated excess conductivity in all the bridges scales as 1/w which may suggest either the presence of a highly conducting region that is dominating the electric transport, or a change in the vortex dynamics in narrow enough bridges. We argue that this effect can only be observed in materials with sufficiently weak vortex pinning.*

It is generally accepted that the superconductivity in superconducting bridges can be suppressed by reducing their dimensions. While sufficiently thick and wide bridges reflect the properties of the bulk material, wide strips with a reduced thickness $d \lesssim \xi$ (where $\xi$ is the Ginzburg-Landau coherence length) can be viewed as quasi two-dimensional [1]. Their properties are then strongly influenced by the thickness $d$, with a certain reduction of the transition temperature to a zero-resistance state [2-5]. Upon further narrowing a bridge down towards to the one-dimensional (1D) limit $w \lesssim \xi$, the critical temperature $T_c$ decreases exponentially with the inverse of the cross section [6], leading to a transition to an insulating state [6-10].

Placing a type-II superconducting strip into an external magnetic field, magnetic-field-induced vortices can exist as long as $w \gtrsim 2\xi$ [11]. In very thin films, vortices can interact in a different way than in their bulk peers, namely via their stray fields in the surrounding space. The characteristic length scale for this interaction is given by the Pearl length $\Lambda = 2\lambda_L^2/d$, which can be substantially larger than the London penetration depth $\lambda_L$ [12]. In wide bridges, where the bridge width $w$ is larger than all length scales that are relevant for superconductivity, the vortex interactions are long-range logarithmic as a function of vortex separation $r$ for $r < \Lambda$, and they determine the superconducting properties in clean enough samples. It has been suggested that in narrow bridges $w < \Lambda$, the interaction becomes short-range exponential for $r > w/\pi$ [13], thereby excluding a Berezinskii-Kosterlitz-Thouless (BKT) transition [14,15]. While surface barriers for vortex entry also play an important role in this low-field limit [16], they are negligible in the high-field limit $B \sim B_{c2}$. In this article, we study the transition to superconductivity for amorphous superconducting films in this high-field limit, as a function of the bridge width $w$ for $w < \Lambda$ and $w > \Lambda$.

We have fabricated micro-bridges based on four amorphous WSi and MoSi films of various thickness, with ten different bridge widths ranging from 2 μm to 1000 μm (fabrication details are provided in the Supplemental Material S1 [17]), and performed detailed transport measurements on them. Figure 1 shows the respective resistive transitions to the superconducting states in magnetic fields up to $B = 5$ T perpendicular to the films. To facilitate a comparison, the original resistance data have been converted to the respective sheet resistances $R_s$. In order to eliminate any minor remaining variations in $R_s$ in the normal state due to uncertainties in the geometric dimensions, we normalized the data to the normal-state sheet-resistance values ($R_n$) of the 100-μm-wide bridges at $T = 7$ K ($T = 10$ K for the 6.2-nm-MoSi film) and $B = 0$ T. The bridges prepared from the 100-nm-thick WSi film have a zero-field critical temperature $T_c(0) \approx 4.95$ K, close to the maximum $T_c$ for amorphous WSi [18-21]. The corresponding critical temperature of the 4-nm-thick WSi film is reduced to $T_c(0) \approx 3.42$ K, in agreement with Refs. [18-21]. The 6.2-nm- and 4.5-nm-thin MoSi films show $T_c(0) \approx 6.85$ K and 5.15 K, respectively, which are the highest reported values for MoSi films in this thickness range to the best of our knowledge [22]. The material parameters relevant for superconductivity for these films are tabulated in section S2 in the Supplemental Material [17]. In zero magnetic field ($B = 0$ T), all bridges made from a particular film show the same critical temperature and temperature dependence of $R_S(T)$ because the respective coherence lengths are more than three orders of magnitude smaller than the width of the narrowest 2-μm-wide bridge [17].

With increasing magnetic field, the sheet resistance curves $R_s(B,T)$ are significantly broadened, and the transitions are shifted towards lower temperatures along with the reduction of the respective critical temperatures $T_c(B)$. The $R_s(B,T)$ data of the bridges made from the 100-nm-thick WSi film show a shoulder-like drop with decreasing temperature before zero resistance is reached (see section S3 in the Supplemental Material [17]), which is reminiscent of corresponding features observed at the first-order solidification of a vortex fluid to a vortex lattice in high-temperature superconductors [23]. Its occurrence is a strong indication for the high quality of our films and supports the notion that bulk vortex pinning in amorphous superconducting films is weak enough to allow for the occurrence of this transition [24].

All our data show a further unexpected striking phenomenon: upon lowering the temperature, the $R_s(B,T)$ of the bridges for a particular film and magnetic field separate in such a way that the resistivity in narrow bridges is significantly suppressed, thereby leading to a narrowing of the transition to the zero-resistance state (Fig. 1). For each of the films, this separation occurs at a well-defined, field-dependent temperature $T^*(B)$ that is independent of the bridge width [see Fig. 2(a) as an example]. It coincides with the temperature where the derivatives $dR_s/dT$ for a given film and magnetic field show a sharp maximum [Fig. 2(b)], indicating that there may be a change in the dissipative mechanism for electric-current flow.

To quantify the observed reductions of $R_s(B,T)$ in our experiments, we have determined the temperatures $T_{max}$ where the differences between the resistivities of the 1000 μm and the 2 μm wide films are largest for each film and each magnetic field. In Figs. 3(a)-3(d) we show the evolution of $R_s(T_{max})$ as functions of the bridge width $w$. The observed reduction of the resistivity as a function of $w$ is most prominent below the scale of the Pearl length (with $\Lambda$ between 8 μm and 345 μm respectively [17]), but becomes almost immeasurably small for $w > 500$ μm $> \Lambda$.

We now quantitatively analyze the additional conductivity $\Delta\sigma$ that is associated with the reduction of $R_s(B,T)$ with decreasing $w$. The $R_s(B,T)$ data for $w = 1000$ μm and $w = 500$ μm are hardly distinguishable, and we therefore use the resistivity of the widest bridge ($w = 1000$ μm) of each film as a reference representing the value for an infinite film of equal thickness. We can plot this additional conductivity for a given bridge width $w$ at $T = T_{max}$, $\Delta\sigma(T_{max},w) = \sigma(T_{max},w) - \sigma(T_{max}, w = 1000$ μm) as a function of $w$ for different magnetic fields [see Figs. 3(e)-3(h)]. For these plots we normalized the conductivities to the respective values $R_n$. As a general trend, these additional conductivities scale in all cases almost exactly as $1/w$ over at least 2 decades. We note that similar analyses for $T \neq T_{max}$ yield the same $1/w$-type of scaling of $\Delta\sigma(T,w)$. This peculiar width dependence is compatible with the presence of a localized region along the strip with width-independent size $s < w$ exhibiting a higher conductivity $\sigma_s$ than the rest of the film with conductivity $\sigma_0$. The averaged conductivity would be $\sigma_{av} = \sigma_0 + s(\sigma_s - \sigma_0)/w$, and the excess conductivity $\Delta\sigma = \sigma_{av} - \sigma_0$ scales as $1/w$ for a given film and magnetic field $B$, and vanishes for an infinite film with $\sigma_{av} = \sigma_0$. As long as the spatial variation of $\sigma(x)$ from $\sigma = \sigma_s$ to $\sigma = \sigma_0$ occurs over a short enough length scale $s \ll w$, this result is insensitive to how exactly $\sigma(x)$ varies from $x = 0$ to $x > s$. Any deviations from a $1/w$-scaling for small bridge widths would indicate that $w$ becomes comparable to $s$ with $\sigma_{av} \rightarrow \sigma_s > \sigma_0$ for even smaller values of $w$. As we do not see any deviation from this scaling, the length scale $s$ within this model (which may depend on magnetic field) must be smaller than 2 µm, both in the thick and in the thin films.

We briefly discuss possible scenarios invoking such highly conductive channels in superconducting strips. We first state that the suppression of $R_s(B,T)$ cannot be explained by the presence of conventional surface barriers [16]. It is known that such barriers inhibit dissipative vortex flow transverse to the current and therefore result in a reduction of the resistance. The maximum magnetic field, below which surface barriers can play a role, is in the limit $\kappa \gg 1$ given by $B_s \approx \phi_0/(4\pi\xi\lambda)$ [25], and in superconducting strips by $\phi_0/(2\pi\xi w) < B_s$ for $w \ll \Lambda$ [26]. It amounts to $\approx 73$ mT at most in our case [17] and is probably much smaller close to the critical temperature, where both $\lambda_L$ and $\xi$ diverge. As $B \gg B_s$ in our investigations, with no signs of any weakening of the effect in the limit $B \rightarrow B_{c2}$, we can definitely exclude that conventional surface barriers are responsible for the reduction of $R_s(B,T)$. Most interestingly, a qualitatively similar sharpening of the resistivity in ~ 300-µm-wide strips of $Bi_2Sr_2CaCu_2O_8$ films with thickness $d \approx 13$ µm [27] could be very well explained by the presence of surface barriers. It was reported to be most pronounced around $B \approx 50$ mT but became negligible beyond $B \approx 2$ T. For the respective experiments we estimate with $\lambda_L \approx 190$ nm and $\xi \approx 1$ nm [28] a $B_s \lesssim 1$ T, in favorable agreement with the observation that the effect vanished in large enough magnetic fields [27].

The observation of a width independent critical current in superconducting 200-nm-thick MoGe strips [29] has been explained the as being a consequence of a vortex-free channel-region of size $W'$ at the edges caused by edge barriers. However, those strips appeared to become fully penetrated by the magnetic field beyond a certain limiting field $B' = B_{c2}/2.6$ where $W' < \xi$, and the effect indeed vanished at higher fields. By very contrast, the reduction of resistivity in our experiments extends up to $B_{c2}$. The existence of such vortex-free superconducting channels would also imply that $\sigma_s$ and $\Delta\sigma \to \yen$ and therefore $R = 0$ for small enough transport currents [29]. A large enough current should quench such channels, and the width-dependent effect should disappear. The excess conductivity $\Delta\sigma$ in our case is finite, leading to a finite resistivity $R_S(B,T)$ down to the lowest applied current density $j$ = 0.2 A/cm [17]. We have verified on the 4.5-nm-MoSi bridges that the separation of $R_S(B,T)$ persists over a wide current-density range, and $T^*$ remains unchanged over five decades of $j$ up to $2 \times 10^4$ A/cm (section S4 in the Supplemental Material [17]).

On more general grounds, superconducting channels of size $\approx \xi$ along the boundaries of an ideal superconductor have been proposed to occur in a magnetic-field range exceeding $B_{c2} < B < B_{c3} \approx 1.7\, B_{c2}$ [30]. Their possible existence (with a large normal conducting region as $B > B_{c2}$) cannot explain our results either, however. First, it is the field $B_{c3}$ that would define the boundary to the zero-resistance state and lead to unphysically low $B_{c2} = B_{c3}/1.7$ values in our films. Second, such channels should occur in all type-II superconductors with similar material parameters and geometry for $B > B_{c2}$, also in strongly-pinning superconductors such as NbN, which we can rule out by experiment (see below). Nevertheless, it is conceivable that the dissipation mechanism in the narrow regions near the edges that are flux-free without a transport current [31,32], is different from the bulk in the presence of a transport current, leading to different vortex-flow velocities under the influence of an electric current. As we are not aware of any corresponding quantitative analysis we must leave this question open for the moment.

The fact that there exists a separation of the $R_S(B,T)$ data even in large magnetic fields also calls for a discussion of possible vortex-pinning effects because the observation of a non-linear $I$-$V$ in the mixed state of superconductors is in most cases associated with vortex pinning. The current-voltage ($I$-$V$) characteristics start indeed to deviate from linearity around the characteristic temperature $T^*(B)$ below which the reduction of the resistance sets in. While hardly discernible in the wide bridges, the corresponding data taken on the 2 μm wide 4.5-nm-thin MoSi bridge show that the respective $I$-$V$ curves are linear above $T^* \approx 4.4$ K at $B = 2$ T, and become increasingly non-linear below it (section S5 in the Supplemental Material [17]). As we do not expect any width-dependent change of the pinning properties across the bulk of the film during the photolithographic structuring process, we might expect enhanced pinning in a localized region located along the edges of the bridges, with bridge-width independent properties and high conductivity. Although this scenario is consistent with the $1/w$ scaling of the excess conductivity, our data suggest that such pinning must be very robust against high magnetic fields up to $\approx B_{c2}$ and large current densities [17].

We now argue that the features reported here can only be observed in films of weakly-pinning superconductors. In Fig. 4, we show corresponding $R_S(B,T)$ data taken on a thin film ($d \approx 5$ nm) of NbN, which is known to be a strongly-pinning superconductor [33]. It is obvious that the large reduction of resistivity in narrow bridges as we measured it in the WSi and MoSi films is absent. We believe that if bulk pinning is strong enough, any additional, comparably weak increase of the electrical conductivity is dominated by bulk pinning and therefore becomes unobservable.

While the reduction of $R_S(B,T)$ in our experiments sets in already for $w \approx \Lambda$, a crossover from macroscopic-to-mesoscopic vortex physics is expected to occur only at a much smaller length scale $w \lesssim 100\xi(0)$ [31], which amounts in our case to less than 1 μm << $\Lambda$ (see Supplemental Material S2 [17]). However, the vortex structure and the vortex interactions may be entirely different from those in infinite films already for $w \lesssim \Lambda$. For large vortex distances ($r \gg \Lambda$), the vortex-vortex interaction in narrow bridges has been shown to be exponentially weak instead of logarithmic [13], and a single vortex is even predicted to carry less than one magnetic flux quantum due to the geometric restriction, particularly near the edges [34]. At present, it has, to our knowledge, not yet been considered how such arguments can be transferred to the case of a dense lattice of Pearl vortices with $r < w \ll \Lambda$ and $B \to B_{c2}$, nor what the possible consequences on measurable quantities might be. We can therefore not rule out that for $w < \Lambda$, a modified vortex interaction in narrow enough bridges changes the vortex dynamics, and that peculiarities near the edges of a bridge play important roles.

According to our results presented here, the characteristic temperature $T^*(B)$ is independent of bridge width and current density and marks the onset of non-linearity in the *I-V* characteristics. At temperatures above $T^*(B)$, the electric currents are immune to geometric effects or vortex-pinning mechanisms. It most likely coincides with the appearance of vortices as well defined quantities, which is a manifestation of persistent macroscopic phase coherence. From this perspective, $T^*(B)$ can be interpreted as the intrinsic critical temperature $T_c(B)$ of a given film. In Fig. 2(c) we show, for comparison, the $T^*(B)$ of the 4.5-nm-thin MoSi bridges, together with corresponding temperatures as derived from an $R/R_n = 0.5$ standard that is often used to define a critical temperature $T_c$. A systematic study on $T^*(B)$ and its relation to standard $T_c$ criteria is underway [35].

Finally, we mention that $R_S(B,T)$ data taken on thin superconducting films have been routinely used to explore magnetic-field induced superconductor-to-insulator transitions for $T \to 0$ [36-41]. It is conceivable that the partly contradictory results reported in the literature may be due to the size effect reported here. According to our results, the intrinsic behavior of 2D films can only be captured on wide enough samples with $w \gg \Lambda$.

In conclusion, we have shown that the resistive transitions in thin films of the weakly-pinning amorphous superconductors WSi and MoSi in a magnetic field are strongly dependent on the width of the samples, with a substantial narrowing near the critical temperatures as soon as the bridge widths become much less than the corresponding Pearl lengths. While the $1/w$ scaling of the excess conductivity may suggest the existence of highly conducting narrow channels along the edges of the bridges in magnetic fields up to $B_{c2}$, we cannot exclude that, in the limit of a dense lattice of Pearl vortices and for $w < \Lambda$, other mechanisms are responsible for the observations. Above a well-defined temperature $T^*(B)$, the transport properties of a given film do no longer depend on geometry. The reported effect may not be relevant in thin films of strongly-pinning superconductors, but it is crucial for the interpretation of resistance data taken on thin films of amorphous superconductors in large magnetic fields.

**Acknowledgements**

We thank to Prof. V. G. Kogan for stimulating discussions. H. Liu and D. Zhu are supported by the Schweizerische Nationalfonds zur Förderung der Wissenschaftlichen Forschung (grant No. 20-175554). K. Smirnov thanks the Russian Science Foundation (RSF) Project No. 18-12-00364.

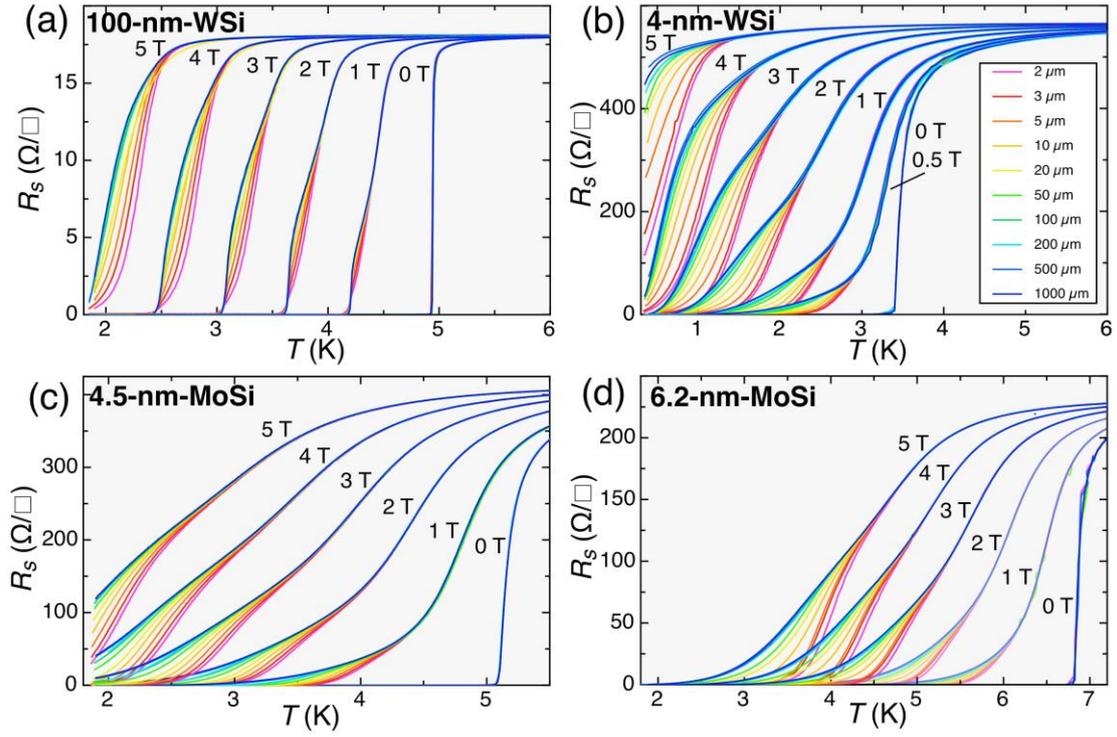

FIG. 1. The sheet resistance $R_s$ as a function of temperature in magnetic fields ranging from $B = 0$ T to $B = 5$ T for 100-nm-thick WSi (a), 4-nm-thick WSi (b), 4.5-nm-thick MoSi (c), and 6.2-nm-thick MoSi (d).

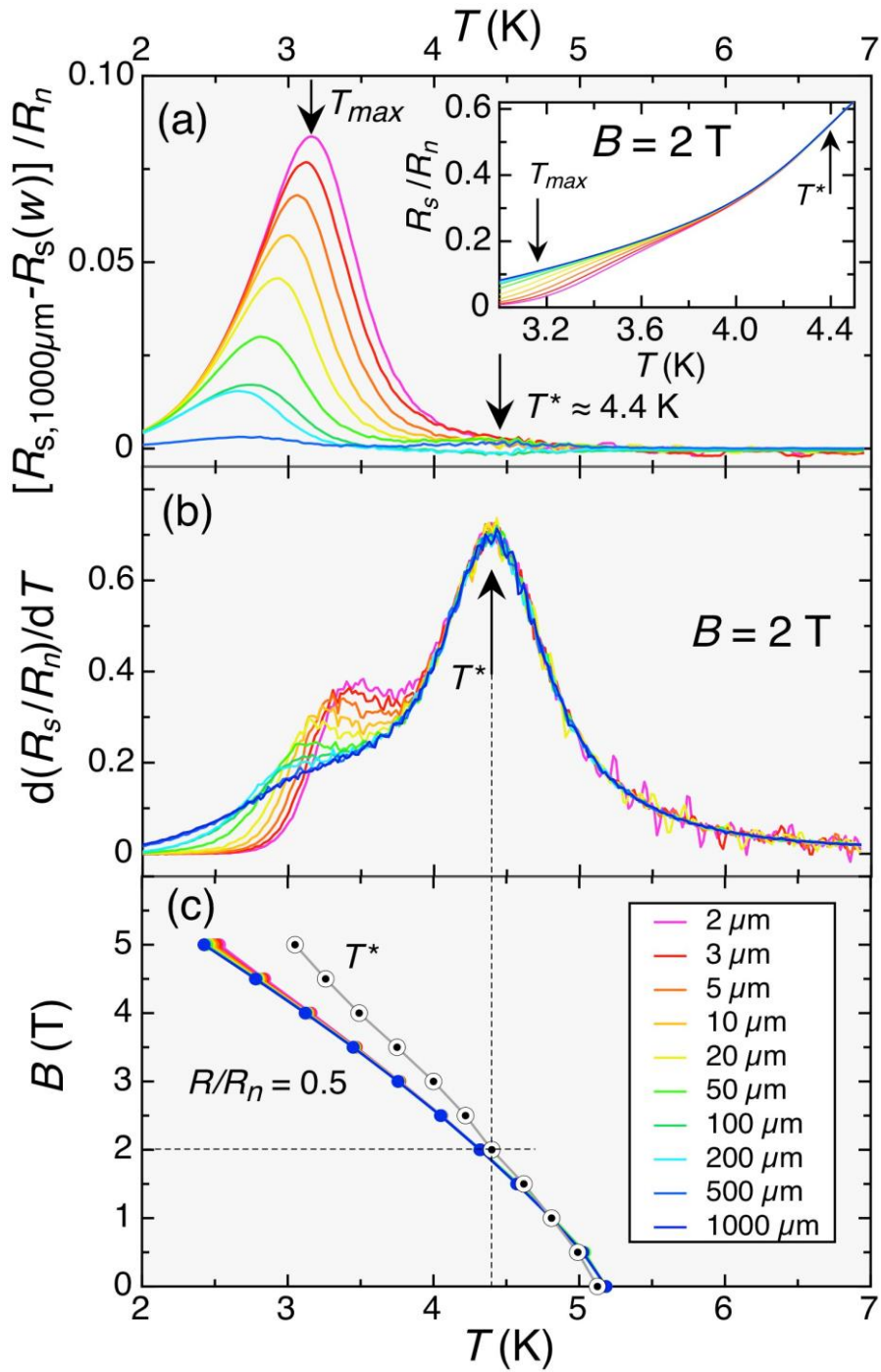

FIG. 2. (a) Difference in the normalized sheet resistance between the 1000 μm wide bridge and the other bridges of the 4.5-nm-thin MoSi film in $B = 2$ T. $T^*$ denotes the temperature around which this difference vanishes and all resistance curves merge (inset), while $T_{max}$ indicates the temperature where the difference between the 1000 μm and the 2 μm data is largest. (b) The derivative of the normalized sheet resistance, showing a sharp maximum at $T^*$. (c) Phase diagram of the same 4.5-nm-thin MoSi bridges, including the temperatures where $R/R_n = 0.5$ and $T^*$. The dashed lines are drawn to guide the eye to the corresponding $T^* \approx 4.4$ K for $B = 2$T as indicated in Figs. 2 (a)-(b).

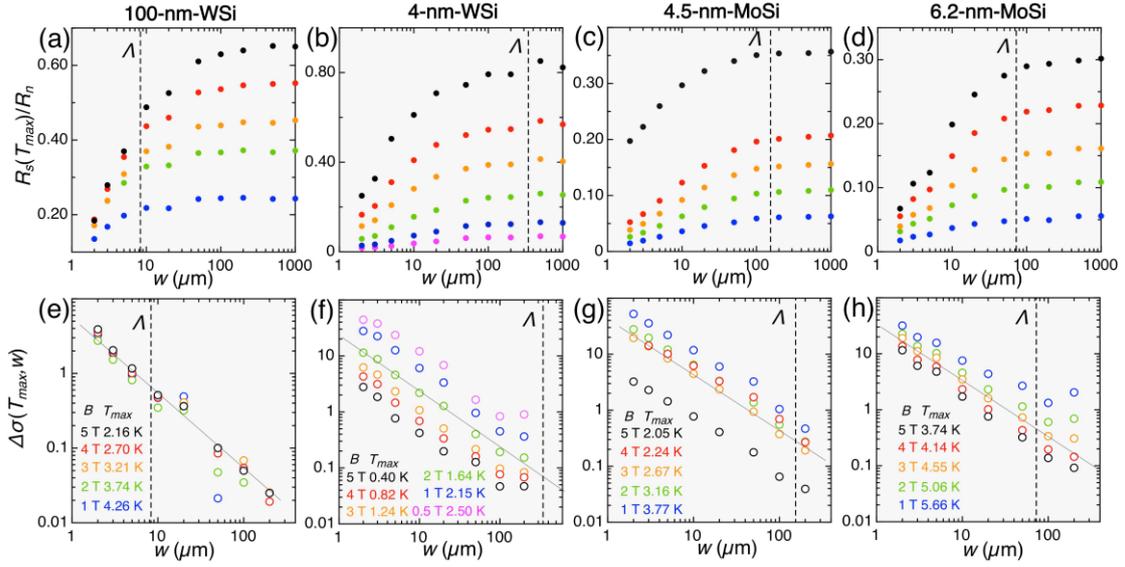

FIG. 3. (a)-(d) Evolution of the normalized sheet resistances as functions of the respective bridge width $w$ in magnetic fields between $B = 1$ T (blue filled circles, lowest curves) and $B = 5$ T (black filled circles, highest curves), in steps of 1 T. The data have been taken at the temperatures $T_{max}$ where the difference between the 1000 μm and the 2 μm data are largest [see Fig. 2(a)]. (e)-(h) Excess normalized conductivities $\Delta\sigma(T_{max}, w)$ relative to the values of the 1000-μm-wide bridges at the respective temperatures $T_{max}$, as functions of the bridge width $w$. The thin solid lines represent a $1/w$ scaling and are to guide the eye. The respective Pearl lengths $\Lambda$ are indicated as vertical dashed lines.

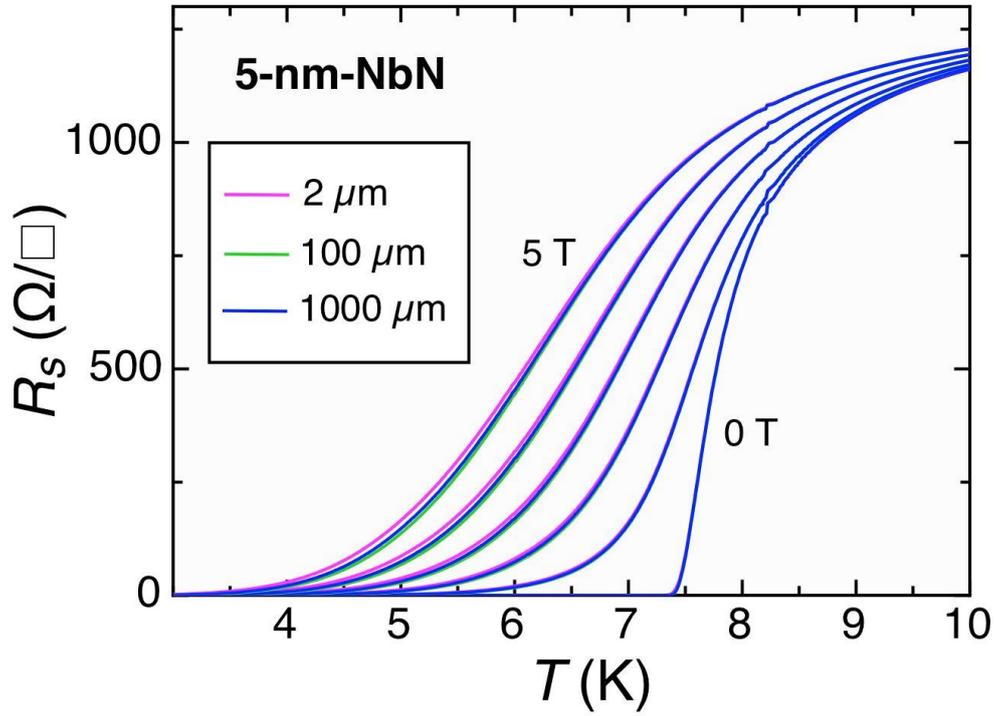

FIG. 4. The sheet resistance $R_s$ as a function of temperature for bridges made from a 5-nm-thick NbN film, in magnetic fields between $B = 0$ T to $B = 5$ T in steps of 1 T. No reduction in resistivity is visible upon decreasing the bridge width from 1000 μm to 2 μm.